\documentclass[12pt]{article}
\usepackage{amssymb,cite}
\usepackage{epsf}

\def\starup#1{\mbox{$\raise1.8ex\hbox{$*$} \kern-.7em#1$}}
\def\krup#1{\mbox{$\raise1.8ex\hbox{$+$} \kern-.7em#1$}}

\begin{document}
\title{Bounds \\ 
       on scalar leptoquark and scalar gluon masses \\
       from S, T, U in the minimal four color symmetry model } 
\author{A.D.~Smirnov\thanks{E-mail: asmirnov@univ.uniyar.ac.ru}\\
{\small\it Division of Theoretical Physics, Department of Physics,}\\
{\small\it Yaroslavl State University, Sovietskaya 14,}\\
{\small\it 150000 Yaroslavl, Russia.}}
\date{}
\maketitle

\vspace{-5mm}
\begin{abstract}
The contributions into radiative correction  
 pa\-ra\-me\-ters S, T, U   
from scalar leptoquark and scalar gluon doublets 
are investigated in the minimal four color symmetry model. 
It is shown that the current experimental data on S, T, U 
allow the scalar leptoquarks and 
the scalar gluons to be relatively light 
(with masses of order of 1 TeV or less), 
the lightest particles are preferred 
to lie below 400 GeV.  
In particular, the lightest scalar leptoquarks with masses 
below 300 GeV 
are shown to be compatible with the current data on 
S, T, U at 
$\chi^2 < 3.1(3.2)$ 
for  
 $m_H=115(300)\,\,GeV$ 
in comparison with 
$\chi^2 = 3.5(5.0)$ 
in the Standard Model.  
The lightest scalar gluon in this case is expected 
to lie below $ 850(720) \,\,\,GeV.$
The possible significance of such particles 
in the t-quark physics at LHC is emphasized. 
\end{abstract}

\newpage

The search for a new physics beyond 
the Standard Model (SM) is  
one of the aims of high energy physics now.  
One of the possible variants
of such new physics can  be the variant induced 
by the possible four color symmetry \cite{PS} 
between quarks and leptons.
This symmetry predicts the vector leptoquarks with 
the lower limit on their masses of order 
$100\,\,TeV$ or less 
\cite{AD1,AD2,RF1,RF2,YF,BL} 
and allows the scalar leptoquarks 
\cite{BVR,HR} 
with the more undefined masses,  
some other colored particles 
are also allowed.  

In addition to the vector leptoquarks 
the four color symmetry with the Higgs mechanism 
of splitting the masses of quarks and leptons 
( MQLS-model, \cite{AD1,AD2} ) predicts  
the scalar leptoquarks and the scalar gluons 
of the doublet structure under the electroweak 
$ SU_L(2) $-group.  
In this approach these doublets are responsible 
for splitting the masses of quarks from those of 
leptons and they are the partners of the standard 
Higgs doublet. 
What can we say about the masses of these scalar 
doublets? The first analysis of the masses of the scalar 
leptoquark doublets by the formalism 
\cite{PT} 
of the radiative 
corrections $ S-, T-,U- $ parameters showed 
\cite{AD3,AD4}   
that some of these particles can be relatively light. 

In the present paper we calculate and discuss the contributions 
into  
$S,\,T,\,U$ 
from both the scalar leptoquark and the scalar gluon doublets,  
accounting the Higgs mechanism of generating the masses 
of these particles from the scalar potential of their 
interactions with the standard Higgs doublet. This is the most 
reasonable way of generating the masses of scalar particles 
and it reduces the region of the fitting parameters of the model 
and gives the possibility to obtain the new bounds 
on the masses of the scalar leptoquarks 
and of the scalar gluons from current experimental data 
\cite{PL,PDG01} 
on $S,\,T,\,U$.

The scalar leptoquark and scalar gluon doublets  
$ S_{a\alpha}^{(\pm)}$ and $F_{ja}$ 
to be discussed here belong to the (15,2,1)-multiplet 
$\Phi_{i,a}^{(3)}$ 
of the 
$SU_V(4) \times SU_L(2)\times U_R(1)$-group 
of the MQLS-model \cite{AD1,AD2} 
with VEV $\eta_3$, 
here    
$ a=1,2$  and $i=1,2\,...\,15$ are 
the $SU_L(2)$ and $SU_V(4)$ indexes and 
$\alpha=1,2,3$ and $j=1,2\,...\,8$ are the indexes of the ordinary 
color $SU_c(3)$ group. 
This multiplet together with (1,2,1)-multiplet 
$\Phi_{a}^{(2)}$ 
with VEV $\eta_2$  
generates the fermion masses by Higgs mechanism 
and splits the masses of quarks from those of leptons.  

Below we consider the scalar leptoquark doublets 
in the case of the 
the simplest scalar leptoquark mixing 
and with neglect of the small parameter 
$\xi^2= \frac23g_4^2\eta_3^2/m_V^2 \ll 1 $ 
of the model. 
In this case the scalar leptoquark doublets can be written as 

\begin{eqnarray}
S^{(+)} = 
\left ( \begin{array}{c}
S_1^{(+)}\\
c\,\, S_1 + s\,\, S_2
\end{array} \right )  ,
S^{(-)} = 
\left ( \begin{array}{c}
S_1^{(-)}\\
-s\,\, \starup{S_1} + c\,\, \starup{S_2}
\end{array} \right )  ,
\label{eq:SpSm}
\end{eqnarray}
where $S_1,\,\,S_2$ are the mass eigen states of 
the scalar leptoquarks with electric charge 2/3 and 
$c=\cos\theta,\,\,s=\sin\theta,\,\,\theta $ 
is the scalar leptoquark mixing angle. 

In the case (\ref{eq:SpSm}) 
the contributions 
$S^{(LQ)}$, $T^{(LQ)}$,$U^{(LQ)}$ 
into 
$S,\,T,\,U$ 
from the scalar leptoquark doublets can be obtained 
by simplifying the general expressions   
of ref.\cite{AD3} 
and take the form 

\begin{eqnarray}
S^{(LQ)} &=& \frac {n_c}{12 \pi} \Bigg \{
-Y_+^{SM} 
\Big \lbrack
c^2 \ln\frac {m_{+}^2}{m_1^2}+s^2 \ln\frac {m_+^2}{m_2^2} 
\Big \rbrack
\label{eq:slq}\\
&&-Y_-^{SM} 
\Big \lbrack
s^2 \ln\frac {m_-^2}{m_1^2}+c^2 \ln\frac {m_-^2}{m_2^2}
\Big \rbrack
+ 4c^2s^2f_2(m_1,m_2)\Bigg\}, 
 \nonumber \\
T^{(LQ)} &=& \frac {n_c}{16 \pi s_W^2c_W^2m_Z^2} \Bigg \{ 
c^2
\Big \lbrack
f_1(m_{+},m_1)+f_1(m_{-},m_2)  
\Big \rbrack
\label{eq:tlq}\\
&&+ s^2
\Big \lbrack
f_1(m_{+},m_2)+f_1(m_{-},m_1)  
\Big \rbrack
- 4c^2s^2f_1(m_1,m_2)\Bigg\}, 
 \nonumber \\
U^{(LQ)} &=& \frac {n_c}{12 \pi} \Bigg \{ 
c^2
\Big \lbrack
f_2(m_{+},m_1)+f_2(m_{-},m_2)  
\Big \rbrack
\label{eq:ulq}\\
&&+ s^2
\Big \lbrack
f_2(m_{+},m_2)+f_2(m_{-},m_1)  
\Big \rbrack
- 4c^2s^2f_2(m_1,m_2)\Bigg\}, 
 \nonumber
\end{eqnarray}
where 
\begin{eqnarray}
f_1(m_1,m_2) & = & {m_1}^2 + {m_2}^2 - \frac {2m_1^2m_2^2}{{m_1}^2 - {m_2}^2}
\ln{ \frac {m_1^2}{m_2^2}}\,, 
\label{eq:f1} \\
\nonumber\\
f_2(m_1,m_2) & = & - \frac {5{m_1}^4 + 5{m_2}^4 - 22{m_1}^2{m_2}^2 }
{3({m_1}^2 - {m_2}^2)^2}
\nonumber \\
& + & \frac {{m_1}^6 - 3{m_1}^4{m_2}^2 - 3{m_1}^2{m_2}^4 + {m_2}^6}
{({m_1}^2 - {m_2}^2)^3}
\ln{ \frac {m_1^2}{m_2^2}}\,,
\label{eq:f2}
\end{eqnarray}
$n_c=3,\,\,  
Y_\pm^{SM}=1\pm 4/3,\,\,  
m_{+}=m_{S^{(+)}_1}=m_{5/3},\,\,  
m_{-}=m_{S^{(-)}_1}=m_{1/3}, \,\,
m_{1,2}=m_{S_1,S_2}=m_{2/3,2/3'}.$  
Here indexes 5/3, 1/3, 2/3 of the masses denote the electric charges 
of the corresponding scalar leptoquarks. 
Notice that the contributions  
 $T^{(LQ)}$ and $U^{(LQ)}$ 
from the scalar leptoquark doublets are not positive definite 
due to the $S_1-S_2$- mixing 
and can be negative if 
$ m_{+}, m_{-} $  
are between $m_1$ and $m_2$. 

The scalar gluon doublets 
can be written as  
\begin{eqnarray}
F_j = 
\left ( \begin{array}{c}
F_{1j}\\
(\phi_{1j}+i\phi_{2j})/\sqrt{2}
\end{array} \right )  ,
\label{eq:Fj}
\end{eqnarray}
where the charged fields $F_{1j}$ and the neutral fields 
$\phi_{1j},\phi_{2j},j=1,2,...,8$ are the mass eigen state fields 
(in general case the real and imaginary parts $\phi_{1j},\phi_{2j}$ 
of the down component of the doublet $F_j$ can be splitted in the mass).  
The scalar gluon doublets (\ref{eq:Fj}) 
give the contributions into 
$S$, $T$, $U$ 
of the form 
\begin{eqnarray}
\hspace{-5mm}S^{(F)}&=&-\frac {k_F}{24 \pi} 
\Bigg\{ 
\ln{ \frac {m_{F_1}^2}{m_{\phi_1}^2}}+ 
\ln{ \frac {m_{F_1}^2}{m_{\phi_2}^2}}- 
f_2(m_{\phi_1},m_{\phi_2}) 
\Bigg\},   \label{eq:sf} \\
\hspace{-5mm}T^{(F)}&=&\frac {k_F}{32\pi{c_W}^2{s_W}^2{m_Z}^2} 
\Bigg\{ 
f_1(m_{F_1},m_{\phi_1})+
f_1(m_{F_1},m_{\phi_2})- 
f_1(m_{\phi_1},m_{\phi_2}) 
\Bigg\},
\label{eq:tf}  \\ 
\hspace{-5mm}U^{(F)}&=&\frac {k_F}{24 \pi} 
\Bigg\{ 
f_2(m_{F_1},m_{\phi_1})+
f_2(m_{F_1},m_{\phi_2})- 
f_2(m_{\phi_1},m_{\phi_2}) 
\Bigg\},
\label{eq:uf} 
 \end{eqnarray}
where $k_F=8 $ and $f_1(m_1,m_2)$ and $f_2(m_1,m_2)$ are defined by 
eqs.(\ref{eq:f1}), (\ref{eq:f2}).
The contributions $T^{(F)}$ and $U^{(F)}$ are not positive definite 
and they are negative if $m_{F_1}$ is between $m_{\phi_1}$ and 
$m_{\phi_2}$. 

Let the masses of the scalar leptoquark and scalar gluon doublets 
are generated from the scalar potential of their interactions with 
the standard Higgs doublet   
by the Higgs mechanism of the symmetry breaking. 
In general case the terms of the scalar potential contributing 
into the scalar leptoquark and scalar gluon masses can be written as 
\begin{eqnarray}
V(\Phi^{(SM)}, S) = \sum_{+,-} \bigg \lbrack 
m_{\pm}^{(0)2}(\krup{S^{(\pm)}}S^{(\pm)}) + 
\beta_{\pm}(\krup{\Phi^{(SM)}}\Phi^{(SM)})(\krup{S^{(\pm)}}S^{(\pm)}) + 
\label{eq:VPhiS}\\
\gamma_{\pm}(\krup{\Phi^{(SM)}}S^{(\pm)})(\krup{S^{(\pm)}}\Phi^{(SM)})  
\bigg \rbrack +  
\bigg \lbrack 
\delta_S \,(\krup{\Phi^{(SM)}}S^{(+)})(\krup{\Phi^{(SM)}}S^{(-)}) + h.c.
\bigg \rbrack ,  
 \nonumber 
\end{eqnarray}

\begin{eqnarray}
V(\Phi^{(SM)}, F) = 
m_{F}^{(0)2}\sum_{j}(\krup{F_j}F_j) + 
\beta_{F}(\krup{\Phi^{(SM)}}\Phi^{(SM)})\sum_{j}(\krup{F_j}F_j) + 
\label{eq:VPhiF}\\
\gamma_{F}\sum_{j}(\krup{\Phi^{(SM)}}F_j)(\krup{F_j}\Phi^{(SM)}) +
\bigg \lbrack 
\delta_F \,\sum_{j}(\krup{\Phi^{(SM)}}F_j)(\krup{\Phi^{(SM)}}F_j) + h.c.
\bigg \rbrack ,  
 \nonumber 
\end{eqnarray}
where $m_{F}^{(0)2}$ and $m_{\pm}^{(0)2}$ are 
the parameters of squared mass dimension and 
$ \beta_{\pm},\,$ $\gamma_{\pm},\,$ $\delta_S,\,$
 $\beta_{F},\,$ $\gamma_{F},\,$ $\delta_F $ are 
the dimensionless coupling constants. 
After symmetry breaking the potentials 
(\ref{eq:VPhiS}),  
(\ref{eq:VPhiF}) 
give the mass matrix of the down scalar leptoquarks 
$ ( S_2^{(+)},\starup{S_2^{(-)}} ) $

\begin{eqnarray}
M = 
\left ( \begin{array}{cc}
M_{11} & M_{12}\\
M_{21} & M_{22}
\end{array} \right ) 
= 
\left ( \begin{array}{cc}
m_{+}^2 + \gamma_{+} \eta^2/2 & \starup{\delta_S}\, \eta^2/2\\
\delta_S \,\eta^2/2               & m_{-}^2 + \gamma_{-} \eta^2/2 
\end{array} \right ) 
\label{eq:MM}
\end{eqnarray}
and the relations for the masses of the scalar gluons  
\begin{eqnarray}
m_{\phi_1,\phi_2}^2 &=& m_{F_1}^2 +
 \gamma_F \eta^2/2 \pm \delta_F \eta^2,  
\label{eq:mphikw}  
\end{eqnarray}
where $\eta= \sqrt{\eta_2^2+\eta_3^2}$ 
is the Standard Model VEV.  

For the case of real $\delta_S$ from the mass matrix (\ref{eq:MM}) 
we have the masses of the scalar leptoquarks with electric 
charge 2/3 and mixing angle in the form 
\begin{eqnarray}
m^2_{1,2}&=&M_{11,22}\mp 
\bigg \lbrack 2M_{12}\cos\theta \sin\theta+(M_{11}-M_{22})\sin^2\theta \,  
 \bigg \rbrack , 
\label{eq:mkw} \\ 
\tan{2 \theta} &=& - 2M_{12}/(M_{11}-M_{22}). 
\label{eq:theta}  
\end{eqnarray}

For the stability of the vacuum the coupling constants 
in the scalar potential are supposed below to satisfy 
some conditions ensuring the positiveness of the total 
scalar potential   
\begin{eqnarray}
V(\Phi^{(SM)}, S) + V(\Phi^{(SM)}, F) +  
\lambda_{SM}(\krup{\Phi^{(SM)}}\Phi^{(SM)})^2 +  
 \nonumber \\ 
\sum_{+,-} \lambda_{\pm}(\krup{S^{(\pm)}}S^{(\pm)})^2 + 
\lambda_F (\sum_{j}(\krup{F_j}F_j))^2 > 0.  
\label{eq:Vg0}
\end{eqnarray}

Below we regard the parameters 
\begin{eqnarray}
 m_{1}, m_{2}, m_{\phi_2}, \gamma_{+}, \gamma_{-}, \delta_S, 
\gamma_F, \delta_F  
\label{eq:par}
\end{eqnarray}
as the fitting parameters and  find the masses 
$m_{+}, m_{-}, m_{\phi_1}, m_{F_1} $  
and the mixing angle from 
(\ref{eq:MM}) - (\ref{eq:theta}) 
and then calculate the contributions 
(\ref{eq:slq}) - (\ref{eq:ulq}),  
(\ref{eq:sf}) - (\ref{eq:uf}) 
of the scalar doublets 
into $S,\,T,\,U$. Notice that for validity 
of the perturbation theory the coupling constants 
in the potentials 
(\ref{eq:VPhiS}),  
(\ref{eq:VPhiF}), 
(\ref{eq:Vg0}) 
cannot be too large and this circumstance bounds the allowed 
region of the fitting masses and mixing angle in the formulas 
(\ref{eq:slq}) - (\ref{eq:ulq}),  
(\ref{eq:sf}) - (\ref{eq:uf}). 
We suppose below that all the coupling constants in 
(\ref{eq:VPhiS}),  
(\ref{eq:VPhiF}), 
(\ref{eq:Vg0}) 
do not exceed some maximal value 
$\lambda_{max}$ 
ensuring the validity of perturbation theory.
In the further numerical analysis we restrict ourselves
by the values of  
$\lambda_{max} $
from the region
$\lambda_{max} = 1.0 - 4.0 $ 
which give the reasonable values of the perturbation 
theory expansion parameter of order 
$\lambda_{max}/4\pi = 0.1 - 0.3 $. 

We have carried out the numerical analysis of the contributions 
(\ref{eq:slq}) - (\ref{eq:ulq}), 
(\ref{eq:sf}) - (\ref{eq:uf}) 
using the current experimental values of $ S,\,T,\,U $ 
induced by a new physics  \cite{PL,PDG01} 
\begin{eqnarray}
S_{new}^{exp}&=& -0.03\pm 0.11\,\,(-0.08),
\nonumber\\
T_{new}^{exp}&=& -0.02\pm 0.13\,\,(+0.09),
\label{eq:stue} \\
U_{new}^{exp}&=& \;\;\; 0.24\pm 0.13\,\,(+0.01), 
\nonumber
\end{eqnarray}
where the central values assume $m_H = 115\,\,GeV$ and the change for 
$m_H = 300\,\,GeV$ is shown in parentheses.

Varying the fitting parameters 
(\ref{eq:par}) 
we minimize
$\chi^2$ 
defined as 
$$
\chi^2=\frac{(S-S_{new}^{exp})^2}{(\Delta S)^2}+
\frac{(T-T_{new}^{exp})^2}{(\Delta T)^2}+
\frac{(U-U_{new}^{exp})^2}{(\Delta U)^2},
$$
where 
$S=S^{(LQ)}+S^{(F)}$,  
$T=T^{(LQ)}+T^{(F)}$,   
$U=U^{(LQ)}+U^{(F)}$ 
and
$S^{(LQ)}$, $T^{(LQ)}$, $U^{(LQ)}$ 
and  
$S^{(F)}$, $T^{(F)}$, $U^{(F)}$  
are the contributions 
(\ref{eq:slq}) - (\ref{eq:ulq}) and  
(\ref{eq:sf}) - (\ref{eq:uf}).  
 $\Delta S,\Delta T, \Delta U$ 
are  the experimental errors in 
(\ref{eq:stue}). 

To clear up the possible effect of the scalar leptoquark 
and scalar gluon doublets on   
$S,\,T,\,U$ 
we vary the masses of these particles so that  
\begin{eqnarray}
m_{1}, m_{2}, 
m_{\pm}, 
m_{F_1}, 
m_{\phi_1}, 
m_{\phi_2} \ge m^{lower}_{scalar} ,
\label{eq:mlower}
\end{eqnarray} 
where 
$m^{lower}_{scalar}$ 
is a lower limit on the masses of these particles. 
 After minimization of 
$\chi^2$ 
under condition (\ref{eq:mlower})
we have analysed the dependence of 
$\chi^2_{min}$ 
on this lower limit 
$m^{lower}_{scalar}$ 
and on upper limit  
$\lambda_{max}$ 
on the coupling constants of the scalar potential. 

The Fig.1 and Fig.2 show 
$\chi^2_{min}( m^{lower}_{scalar},\lambda_{max})$ 
as a function of the lower limit 
$ m^{lower}_{scalar} $  
for $m_H=115\,\,GeV$ 
and for $m_H=300\,\,GeV$ 
respectively by the curves $a(b)$ for 
$\lambda_{max}=1.0(4.0)$ 
for the case without scalar leptoquark mixing 
($\theta=0$, this case is slightly preferred by 
$\chi^2$ minimum ). 
The horizontal lines denote  
$\chi^2_{SM} = 3.5 $ 
and  
$\chi^2_{SM} = 5.0 $ 
of the compatibility of the SM zero values of 
$S,\,T,\,U$ 
with the experimental data 
(\ref{eq:stue})  
at $m_H=115\,\,GeV$ 
and $m_H=300\,\,GeV$ 
respectively. 

As seen from the Figs.1,2 the lower limit 
  $ m^{lower}_{scalar} $  
on the masses of the scalar leptoguarks and of the scalar gluons 
is allowed by data  
(\ref{eq:stue})  
to vary within wide limits from high values 
when the contributions from these particles into 
$S,\,T,\,U$ 
are negligibly small to values of order of 
   $1\,\,TeV$ 
or less. 
It is interesting that in both cases the more light 
particles agree with the data 
(\ref{eq:stue})  
even slightly better than in the SM.  
For 
$m_H=115\,\,GeV$ 
(Fig.1) such slight improvement of the agreement 
takes place for 
$ m^{lower}_{scalar} < 400 \,\,\,GeV $  
whereas in the case of 
$m_H=300\,\,GeV$ 
 (Fig.2) such improvement is seen in all the region 
of the lightest masses of order of  
1 TeV or less and it is more appreciable also for 
$ m^{lower}_{scalar} < 400 \,\,\,GeV $. 
In particular the scalar leptoquarks 
with the lightest masses of order of    
$ m^{lower}_{scalar} < 300 \,\,\,GeV $ 
(and for $\lambda_{max}=4.0 $ ) 
are compatible with the data   
(\ref{eq:stue})  
at 
$\chi^2 < 3.1 (3.2) $ 
for 
$m_H=115 (300)\,\,GeV$ 
in comparison with 
$\chi^2_{SM} = 3.5 (5.0) $ 
in the SM. 
The mass of the lightest scalar gluon in this case 
is expected to be 
$ m_{\phi_2} < 850(720) \,\,\,GeV $.

The slight improvement for 
$ 400 \,\,\,GeV \le m^{lower}_{scalar} \le 1 \,\,\,TeV $ 
is caused by the sufficient contributions which are 
given by the scalar leptoquark and scalar gluon doublets 
into $ S $ and $ T $, the contributions into $ U $ 
in this case are negligibly small. 
For the more light particles 
($ m^{lower}_{scalar} < 400 \,\,\,GeV $ ) 
the light scalar leptoquark doublets give the more 
noticeable contribution into $ U $ with the simultaneous 
cancellation of their relatively large contributions 
into $ S $ and $ T $ with those from the scalar gluon 
doublets. 
As a result the more appreciable agreement 
with the data  
(\ref{eq:stue})  
is achieved. 

For example the scalar leptoquarks and gluons 
with the masses
\begin{eqnarray}
\nonumber
m_{5/3}=330\, GeV,\,\,\,\, 
m_{1/3}=430\, GeV,\,\,\,\,
m_{F_1}=850\, GeV, \hspace{35mm} \\ 
m_{2/3}=250\, GeV,\,\,\,\, 
m_{2/3}'=250\, GeV,\,\,\,\,
m_{\phi_1}=1040\, GeV,\,\,\,\,
m_{\phi_2}=770\, GeV,\,\,\,\,
\label{eq:mlf1}
\end{eqnarray}
give the contributions
\begin{eqnarray}
\nonumber
S^{(LQ)}&=& -0.07,\,\,\,\,
T^{(LQ)}=  2.03,\,\,\,\, 
U^{(LQ)}= 0.02, \\
\nonumber
S^{(F)}&=&  0.03,\,\,\,\,
T^{(F)}= -2.05,\,\,\,\, 
U^{(F)}=-3 \cdot 10^{-3},   \\
\nonumber
S&=& -0.04,\,\,\,\,
T= -0.02,\,\,\,\, 
U= 0.02 ,
\end{eqnarray}
which agree with (\ref{eq:stue}) for  
$m_H = 115\,\,GeV $ with $\chi^2= 2.9 $  
(in comparison with $\chi^2= 3.5 $ of the SM). 

In a similar way for the  masses 
\begin{eqnarray}
\nonumber
m_{5/3}=430\, GeV,\,\,\,\, 
m_{1/3}=430\, GeV,\,\,\,\,
m_{F_1}=650\, GeV, \hspace{35mm} \\ 
m_{2/3}=250\, GeV,\,\,\,\, 
m_{2/3}'=250\, GeV,\,\,\,\,
m_{\phi_1}=890\, GeV,\,\,\,\,
m_{\phi_2}=550\, GeV,\,\,\,\,
\label{eq:mlf2}
\end{eqnarray}
we obtain
the contributions  
\begin{eqnarray}
\nonumber
S^{(LQ)}&=& -0.17,\,\,\,\,
T^{(LQ)}=  3.39,\,\,\,\, 
U^{(LQ)}= 0.04,  \\
\nonumber
S^{(F)}&=& 0.05,\,\,\,\,
T^{(F)}= -3.32,\,\,\,\, 
U^{(F)}=-0.01,  \\
\nonumber
S&=& -0.12,\,\,\,\,
T= 0.07,\,\,\,\, 
U= 0.03, 
\end{eqnarray}
which agree with the data 
(\ref{eq:stue}) 
for  
$m_H = 300\,\,GeV $ with $\chi^2= 2.9 $  
(in comparison with $\chi^2= 5.0 $ of the SM). 

The lightest scalar leptoquark masses in 
(\ref{eq:mlf1}), (\ref{eq:mlf2}) 
are compatible with the experimental limits 
resulting from the direct search for the leptoquarks. 
The most stringent of these limits are resulted from 
the pair production and for the scalar leptoquarks 
of the first generation they give 
\cite{PDG01} 
\begin{eqnarray}
m_{LQ} > 225 \,\,GeV, \,\,\,204 \,\,GeV, \,\,\,79 \,\,GeV 
\label{eq:mlqe} 
\end{eqnarray} 
under assuming the branching ratios 
$B(eq)=1, \,\,\, 0.5, \,\,\, 0 $ 
respectively. 

It should be noted that in the model under consideration 
the coupling constants of the scalar leptoquark doublets 
with the fermions 
( and those of the scalar gluon doublets )
are proportional to the ratios of the 
fermion masses to the SM VEV 
$ \eta=246 \,\,GeV $ 
( the general form of this interaction can be found in 
ref.\cite{PovSm1} )
and for ordinary quarks these coupling constants are small. 
The dominant decay modes of such leptoquarks are the 
modes with heavy quarks ( predominantly with t-quark ) 
whereas the branching ratio for the first generation 
is small 
$ 0 < B(eq) \ll 0.5 $. 
So the lower experimental limit on the masses of such 
scalar leptoquarks can be near the lowest value in 
(\ref{eq:mlqe}), 
the masses $ m_{1/3}, \,\, m_{5/3} $ 
of the other scalar leptoquarks are in this case 
compatible with other experimental limits 
( including those for the second and for 
the third generations ) 
resulting from the direct search for leptoquarks. 

It should be noted also that the light scalar leptoquarks 
can be also compatible with the indirect leptoquark mass limits 
resulting from the rare decays of 
$K^0_L \rightarrow \mu e$ 
type. 
Due to the smallness of the coupling constants 
of the scalar leptoquark interaction 
with d- and s- quarks 
the contributions of the scalar leptoquarks into 
$K^0_L \rightarrow \mu e$ 
width 
can be sufficiently small to satisfy the stringent 
experimental limit 
$Br(K^0_L \rightarrow \mu e) < 4.7 \cdot 10^{-12}$ 
 \cite{PDG01} 
on the branching ratio of this decay,  
even for the relatively light masses of the scalar leptoquarks. 
   
Thus, the current direct and indirect mass limits 
for leptoquarks do not exclude the relative light scalar 
leptoquark doublets considered here whereas the experimental 
data on 
$ S, \,\, T, \,\, U $ 
not only allow the existence of such particles 
but even slightly prefer them 
to have the masses of order of 
$ m^{lower}_{scalar} < 400 \,\,\,GeV $. 
The search for such scalar leptoquarks and scalar gluons 
in the processes with heavy quarks 
 ( predominantly with t-quark ) 
at LHC is of interest.  

It should be noted that the presence of the so light 
new particles can also affect the new physics 
at high energies. In particular these particles 
can affect the gauge coupling constant unufication 
in GUT approaches. As known the SM without any new 
physics up to the GUT mass scale 
$M_{GUT}$  
("grand desert") do not unify three coupling constants 
at any mass scale. But such a unification can be 
possible if an intermediate new physics below   
$M_{GUT}$  
(such as the four color symmetry physics with mass scale 
$M_c$) is assumed. For example in the model under 
consideration in the case of the scalar sector 
containing, for simplicity, only the standard Higgs doublet 
and the (4,1,1) multiplet with VEV 
$\eta_1 \sim M_c \sim 10^{11} \div 10^{12} \,\, GeV$ 
all three coupling constants do converge in one point at 
$M_{GUT} \sim 10^{14} \div 10^{15} \,\, GeV$ 
with 
$\alpha_3(M_{GUT}) = \alpha_2(M_{GUT}) = \alpha_1(M_{GUT}) 
\equiv  \alpha_{GUT} \sim 0.023.  $ 
The account of the scalar leptoquark and scalar gluon 
doublets gives the additional contributions 
$\Delta b_3 = 8/3, \,\, \Delta b_2 = 7/3, \,\, \Delta b_1 =37/15  $  
into the factors $b_i$ which define the mass 
scale  evolution of the running coupling constants 
$\alpha_i(\mu)$ through the beta functions $\beta_i$
of the one loop approximation, $ i = 3, 2, 1.$ 
These contributions together with the SM factors    
$ b_3^{SM} =-7, \,\,  b_2^{SM} = -19/6,  \,\, b_1^{SM} =41/10  $ 
and with the corresponding contributions 
(for $M_c < \mu < M_{GUT} $) 
from the (4,1,1) scalar and from the vector leptoqurks 
determine the mass scale evolution of $\alpha_i(\mu)$ 
from $\mu \sim 1 \,\, TeV$ to $\mu \sim M_{GUT}$. 
The analysys shows that in this case
all three coupling constants  
 $\alpha_i(\mu)$ 
do also converge in one point if
$M_c \sim 10^{10} \div 10^{11} \,\, GeV$  
and 
$M_{GUT} \sim 10^{14} \div 10^{15} \,\, GeV$ 
with
$ \alpha_{GUT} \sim 0.029  $. 
As seen the presence of the relatively light 
scalar leptoquark and scalar gluon doublets
(with mases below 1 TeV)
lowers the four color symmetry 
mass scale $M_c$  and increases the value of 
the unified coupling constant $\alpha_{GUT}$, 
leaving the GUT mass scale $M_{GUT}$ practically 
unchanged.  

As mentioned above the scalar multiplets (1,2,1,) and (15,2,1) 
were introduced to give 
the Dirac masses to the fermions and to split the masses 
of the quarks from those of the leptons 
by the Higgs mechanism. 
The general form of Yukawa interaction of these 
doublets with the fermions makes the fermion masses 
to be arbitrary as they 
are in the SM. The lightness of neutrinos can be 
ensured by the smallness of their Dirac masses 
due to the smallness of the corresponding Yukawa 
coupling constants or, more naturally, 
by the smallness of their Majorana masses 
due to the seesaw mechanism. In the latter case 
the necessary large Majorana mass term of the 
right neutrinos can be 
generated by the interaction of the right fermions 
with an additional $(\overline{10},1,-2)$ scalar 
multiplet with the large VEV of order of $M_c$. 

In conclusion we resume the results of the work. 

 The contributions into radiative correction  
S-, T- ,U- pa\-ra\-me\-ters  
from the scalar leptoquark and scalar gluon doublets 
are investigated 
in the minimal model with 
the four color symmetry,  
accounting the Higgs 
mechanism of generating the masses of these particles. 
It is shown that the current experimental data on S, T, U 
allow the existence of the relatively light scalar 
leptoquarks and scalar gluons 
(with masses of order of 1 TeV or less),  
the more light particles (with masses below 400 GeV) 
are preferred and 
agree with these data better than in 
the Standard Model. 

In particular the scalar leptoquarks   
with the masses of order of 
  $ m^{lower}_{scalar} < 300 \,\,\,GeV $  
are shown to be compatible with current data on 
S, T, U for 
$m_H = 115(300)\,\,GeV $ 
with $\chi^2 < 3.1(3.2) $  
(in comparison with $\chi^2= 3.5(5.0) $ of the SM). 
The lightest scalar gluon in this case is expected 
to lie below $ 850(720) \,\,\,GeV. $

We emphasize the possible significance of such particles
in the top-quark physics at LHC.

\vspace{3mm}
{\bf Acknowledgments}
\vspace{3mm}

The work was partially supported by the Programme 
"Universities of Russia -- Basic Research" 
of Ministry of Education under grant 02.01.25 
and by the Russian Foundation for Basic Research 
under grant 00-02-17883. 
I thank A.V.~Povarov for the help in preparing 
the computer programme of numerical calculations.

\newpage

{\Large\bf Figure captions}

\bigskip

\begin{quotation}

\noindent 
Fig. 1. $\chi^2_{min}( m^{lower}_{scalar},\lambda_{max})$ 
        as a function of the lower limit 
        $ m^{lower}_{scalar} $  
        on the masses of the scalar particles for 
        $m_H=115\,\,GeV$ 
        at $\lambda_{max}=1.0 (a) $ and  
        $\lambda_{max}=4.0 (b) $.  \\

\noindent 
Fig. 2. $\chi^2_{min}( m^{lower}_{scalar},\lambda_{max})$ 
        as a function of the lower limit 
        $ m^{lower}_{scalar} $  
        on the masses of the scalar particles for 
        $m_H=300\,\,GeV$ 
        at $\lambda_{max}=1.0 (a) $ and  
        $\lambda_{max}=4.0 (b) $.  \\ 

\end{quotation}

\newpage
\begin{figure}[htb]
\epsffile[110 450 0 700]{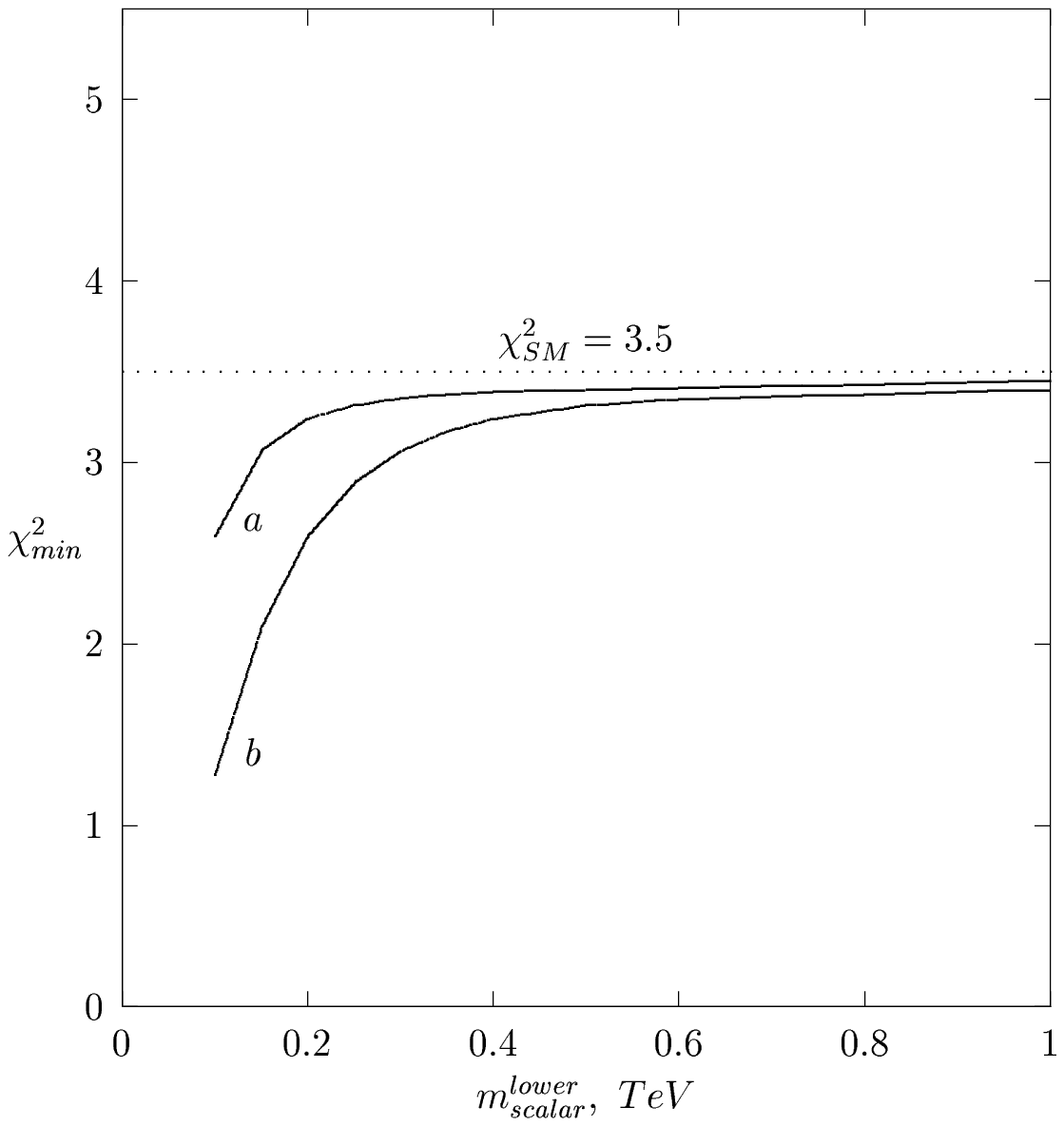}
\end{figure}

\vfill
\centerline{A.D.~Smirnov, Physics Letters B}

\centerline{Fig. 1}

\newpage
\begin{figure}[htb]
\epsffile[110 450 0 700]{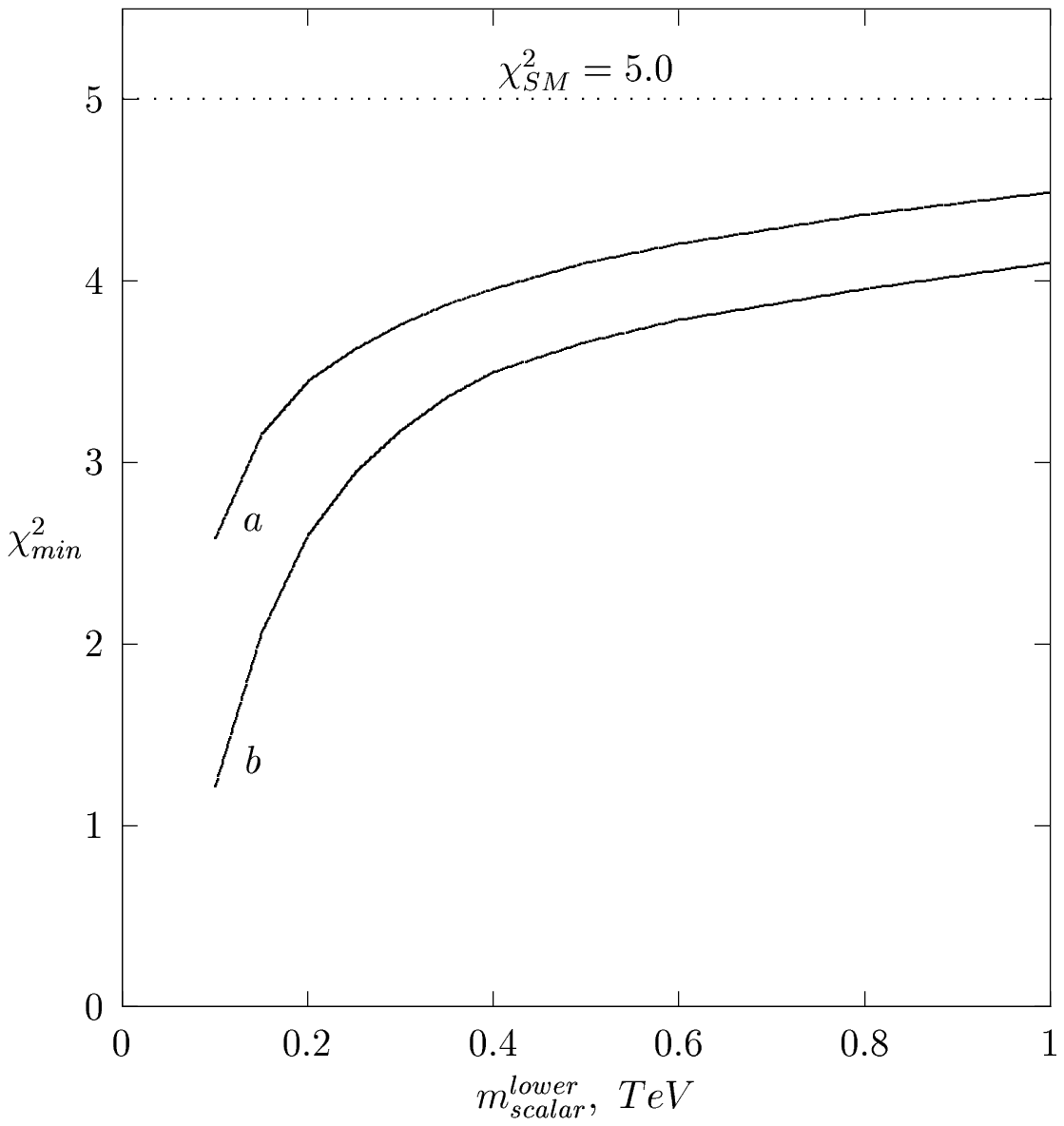}
\end{figure}

\vfill
\centerline{A.D.~Smirnov, Physics Letters B}

\centerline{Fig. 2}


\newpage
\vspace{-5mm}
\begin{thebibliography}{99}
\vspace{-2mm}
\bibitem{PS}
   J.C.~Pati and A.~Salam, Phys.~Rev. D10~(1974)~275.
\bibitem{AD1}
   A.D.~Smirnov, Phys.~Lett. B346~(1995)~297.
\bibitem{AD2}
   A.D.~Smirnov, Yad.~Fiz. 58~(1995)~2252,
                 Phys. At. Nucl. 58~(1995)~2137.
\bibitem{RF1}
   R.~Foot, Phys.~Lett. B420~(1998)~333.
\bibitem{RF2}
   R.~Foot, G.~Filewood, Phys. Rev.~D60~(1999)~115002.  
\bibitem{YF}
   T.L.~Yoon, R.~Foot, hep-ph/0105101, Phys.~Rev. D65~(2002)~015002.
\bibitem{BL}
   A.~Blumhofer, B.~Lampe, Eur.~Phys.~J. C7~(1999),~141. 
\bibitem{BVR}
   W.~Buchm\"uller,R.~R\"uckl, D.~Wyler, Phys.~Lett. B191~(1987)~442. 
\bibitem{HR}
   J.L.~Hewett, T.G.~Rizzo, Phys.~Rev. D56~(1997)~5709.
\bibitem{PT}
   M.E.~Peskin and T.~Takeuchi, Phys.~Rev. D46~(1992)~381.
\bibitem{AD3}
   A.D.~Smirnov, Phys.~Lett. B431~(1998)~119.
\bibitem{AD4}
   A.D.~Smirnov, Yad.~Fiz. 64~(2001)~367,
                 Phys. At. Nucl. 64~(2001)~318.
\bibitem{PL}
   P.Langacker, hep-ph/0110129.
\bibitem{PDG01}
   Particle Data Group, D.E.~Groom et~al., Eur.~Phys.~J. C15~(2000),~1 
   and 2001 partial update for edition 2002 (URL:http://pdg.lbl.gov).  
\bibitem{PovSm1}
   A.V.~Povarov, A.D.Smirnov, Yad.Fiz. 64~(2001)~78,   
                 Phys. At. Nucl. 64~(2001)~74.


\end{thebibliography}
\end{document}